\documentclass[aps,preprint]{revtex4}%
\usepackage{amsfonts}
\usepackage{amsmath}
\usepackage{amssymb}
\usepackage{graphicx}%
\begin{document}
\preprint{IASa001}
\title[Cosmology of scalar field]{Cosmology of an asymptotically free scalar field with spontaneous symmetry breaking}
\author{Kerson Huang$^{1,2}$}
\author{Hwee-Boon Low$^{2}$}
\author{Roh-Suan Tung$^{2}$}
\affiliation{$^{1}$Physics Department, Massachusetts Institute of Technology, Cambridge,
MA, USA 02139}
\affiliation{$^{2}$Institute of Advanced Studies, Nanyang Technological University,
Singapore 639673}
\keywords{commology, dark energy, scalar field, asymptotic freedom}
\pacs{98.80.Bp, 98.80.-k, 03.70.+k, 98.80.Cq}

\begin{abstract}
We solve Einstein's equation with Robertson-Walker metric as an initial-value
problem, using as the source of gravity a Halpern-Huang real scalar field,
which was derived from renormalization-group analysis, with a potential that
exhibits asymptotic freedom and spontaneous symmetry breaking. Both properties
are crucial to the formulation of the problem. Numerical solutions show that
the universe expands at an accelerated rate, with the radius increasing like
the exponential of a power of the time. This is relevant to "dark energy" and
"cosmic inflation". Extension to the complex scalar field will make the
universe a superfluid. The vortex dynamics that emerges offers explanations
for other cosmological problems, namely, matter creation, galactic voids, and
the \textquotedblleft dark mass\textquotedblright.

\end{abstract}
\volumeyear{year}
\volumenumber{number}
\issuenumber{number}
\eid{identifier}
\date[Date text]{date}
\received[Received text]{date}

\revised[Revised text]{date}

\accepted[Accepted text]{date}

\published[Published text]{date}

\startpage{101}
\endpage{102}
\maketitle

Scalar fields have been invoked in cosmological models of "dark energy" [1],
and "cosmic inflation" [2]. For dark energy, i.e., accelerating expansion of
the universe, a scalar field is used as an alternative to Einstein's
cosmological constant. In cosmic inflation, one introduces a scalar field with
spontaneous symmetry breaking, i.e., having a potential with a minimum located
at a nonzero value of the field. Initially the universe was placed at the
"false vacuum" of zero field, and it "inflates" while it "rolls down" the
potential towards the true vacuum. It seems appropriate, therefore, to try to
solve Einstein's equation as an initial-value problem, using a scalar field as
the source of gravity. Mathematical consistency requires that the scalar
potential be asymptotically free. That is, it should be zero at the big bang
and grow as the universe expands, instead of having the opposite behavior, as
with the familiar $\phi^{4}$ theory. Spontaneous symmetry breaking is
desirable for physical consistency, because the existence of a nonzero vacuum
field enables us to neglect quantum fluctuations and treat the entire problem
classically. Both requirements are fulfilled in the scalar field found by
Halpern and Huang (HH) [3,4] on the basis of renormalization-group (RG)
analysis. They show that asymptotic freedom uniquely determines the potential
to be a Kummer function, which has exponential behavior at large fields.

We consider an $N$-component $O(N)$-symmetric real scalar quantum field
$\phi_{n}(x),$ with flat-space Lagrangian density $L\left(  x\right)
=\frac{1}{2}\sum_{n}\left(  \partial\phi_{n}\right)  ^{2}-V\left(  \phi
^{2}\right)  $, where $\phi^{2}=\sum_{n}\phi_{n}^{2}$, and $\hbar=c=1.$ In the
quantum field theory, a high-momentum cutoff $\Lambda$ is the only scale in
the system, and all coupling parameters are scaled by appropriate powers of
$\Lambda.$ As $\Lambda$ changes, the coupling parameters "renormalize" to
compensate for the change, such that the theory is preserved, i.e., the
partition function remains invariant. They trace out an RG trajectory in the
parameter space, the space of all possible scalar theories. The limit
$\Lambda\rightarrow\infty$ corresponds to a fixed point, at which the system
is scale-invariant. One such point is the Gaussian fixed point corresponding
to the massless free field, and this is where we would place the system just
before the big bang. At the big bang the length scale begins to increase, and
asymptotic freedom means that the scalar field leaves this fixed point along a
direction tangent to some RG trajectory. The initial direction determines the
precise form of the potential.

The HH potential has the form $V=$ $\Lambda^{4}U_{b}$, where the dimensionless
potential $U_{b}$ is given by
\begin{equation}
U_{b}(z)=c\Lambda^{-b}\left[  M\left(  -2+b/2,N/2,z\right)  -1\right]
\end{equation}
where $z=8\pi^{2}\phi^{2}/\Lambda^{2}$, and $c$ is an arbitrary constant. The
function $M$ is a Kummer function, with power series $M(p,q,z)=1+zp/q+\left(
z^{2}/2!\right)  p\left(  p+1\right)  /q\left(  q+1\right)  +\cdots$, and
asymptotic behavior $z^{p-q}\exp z$. The potential satisfies $\Lambda\partial
U_{b}/\partial\Lambda=-bU_{b}$ in the neighborhood of the fixed point $\left(
\Lambda=\infty\right)  $; it is an eigenfunction of scale transformation with
eigenvalue $b$. The case $b=2$ corresponds to the free field, and $b=0$
corresponds to the $\phi^{4}$theory. Asymptotic freedom corresponds to $b>0$,
and spontaneous symmetry breaking occurs when $b\,<2$; hence we choose
$0<b<2$. This potential is a lowest-order approximation in $\Lambda^{-1}$,
derived in flat space-time. Higher-order and curvature corrections have not
been calculated, but this should be valid in a neighborhood of the big bang.

In the cosmological problem, we use the Robertson-Walker (RW) metric for a
homogeneous space, with scale parameter $a\left(  t\right)  $, where $t$ is
the time. Since this is the only scale in the system, we put $\Lambda=1/a.$
This signifies that spacetime curvature supplies the cutoff \ for the quantum
scalar field. Einstein's equation, together with the scalar-field equation of
motion in the RW metric, leads to the following initial-value problem:%
\begin{align}
\dot{H}  &  =ka^{-2}-4\pi G\left[  \sum_{n}\dot{\phi}_{n}^{2}-a\left(
\partial V/\partial a\right)  /3\right] \nonumber\\
\ddot{\phi}_{n}  &  =-3H\dot{\phi}_{n}-\partial V/\partial\phi_{n}\nonumber\\
X  &  =0
\end{align}
where $H=\dot{a}/a$, with a dot denoting time derivative, is the Hubble
parameter, $k=0,\pm1$ is the curvature parameter, $G$ is the gravitational
constant, and $X=H^{2}+ka^{-2}-(8\pi G/3)\left(  \sum_{n}\dot{\phi}_{n}%
^{2}/2+V\right)  $. The equation $X=0$ is Friedman's equation ($00$ component
of Einstein's equation), which acts as a constraint on initial conditions. The
existence of the the Cauchy problem [5] in general relativity means that
$\dot{X}=0.$ To preserve this when $V$ depends on the cutoff, we have added
the $\partial V/\partial a$ term in the first equation. This is equivalent to
adding a correction term $-a\left(  \partial V/\partial a\right)  /3$ to the
pressure of the scalar field, making the energy density and pressure
$\rho=\sum_{n}\dot{\phi}_{n}^{2}/2+V$, $p=\sum_{n}\dot{\phi}_{n}%
^{2}/2-V-a\left(  \partial V/\partial a\right)  /3$, respectively. A more
basic derivation would necessitate a reformulation of the action principle,
which we leave for the future.

If the constraint $X=0$ were ignored in choosing initial conditions, one would
generally obtain $H\left(  t\right)  \sim$ $H_{\infty}$ asymptotically,
leading to $a\left(  t\right)  \sim\exp\left(  H_{\infty}t\right)  $. This
would have posed a "fine-tuning" problem, for $H_{\infty}$ would be naturally
on the Planck scale, some 60 orders of magnitude larger than present value.
With the constraint $X=0$, one can show that the time-averaged behavior is
$\left\langle H\left(  t\right)  \right\rangle \sim t^{-p}$, corresponding to
$\left\langle a\left(  t\right)  \right\rangle \sim\exp t^{1-p}$, where $p$
depends on the parameters of the potential, and initial conditions. For
$0<p<1$, the universe is in accelerated expansion. Thus, the vacuum field
supplies dark energy, but its relation to the expansion is neither intuitive
nor simple. The power-law decay of $H$ somewhat eases the fine-tuning problem.
We shall discuss comparison with observations, including the deceleration era
[6], in a separate paper.

Fig.1 shows numerical results for a real scalar field with the HH potential
$(N=1)$. They verify the power law as a time-averaged behavior, and give
$p=0.81$. Fig.2 shows results for a superposition of two HH potentials, which
give $p=0.075$. Numerical computations fail after a time, for they tend to
violate the constraint $X=0$, and need to be supplemented by correction procedures.

The problem of cosmic inflation is inseparable from matter creation, which has
not been taken into account. We have looked into this problem with the real
scalar field, and conclude that efficient matter production calls for
extension to a complex scalar field. This work is in progress, and we report
on some qualitative expectations.

A complex scalar field corresponds to $N=2$, and is not qualitatively
different from the case $N=1$ if the components $\phi_{1},\phi_{2}$ are
uniform in space. However, new physics emerges when we use the complex
representation $\phi,\phi^{\ast}$, with $\phi=F\exp\left(  i\sigma\right)  $,
and allow the phase $\sigma$ to vary in space. The scalar field then describes
a superfluid with superfluid velocity $\mathbf{v}=\mathbf{\nabla}\sigma$.
Whenever a fluid flows there will be vortices, and in the superfluid case we
have quantized vortices. Some immediate consequences of the vortex dynamics
are the following.

1. Whenever two quantized vortex lines cross, they could reconnect, as
pictured by Feynman [7]. Friction on the superfluid, from quantum and thermal
fluctuations, can generate large growing vortex rings, which will emulsify
under reconnections and form a vortex tangle [8], with fractal dimension 1.6
[9]. On the vortex lines, reconnection instantaneously creates two cusps,
which spring away from each other with theoretically infinite speed. The
accompanying burst of kinetic energy could create matter in the form of two
opposing jets. The output rate would be Planck energy in Planck time, or
10$^{18}$GeV in 10$^{-43}$s. As the universe expands, the vortex tangle will
eventually decay, but could have created enough matter for nucleogenesis. The
demise of the vortex tangle would mark the end of the inflation era.

2. There will be remnant vortex lines, at much reduced density. Their vortex
core size must be proportional to the scale $a\left(  t\right)  $ in the RW
metric. Inside the core there is no scalar field, and presumably no matter.
Thus, with the expansion of the universe, vortex cores will grow to become the
presently observed voids in galactic space.

3. Reconnection of fat vortex lines in a later universe could create the
observed cosmic jets and gamma-ray bursts.

4. A random potential can pin a superfluid [10]. A rotating galaxy might drag
portions of the superfluid with it, thus acquiring "dark mass".

\bigskip

{\LARGE References}

\noindent\lbrack1] P.J.E. Peebles and R. Bharat, \textit{Rev. Mod. Phys}.,
\textbf{75, }559 (2003) .

\noindent\lbrack2] L.F. Abbott and S.-Y. Pi, \textit{Inflationary Cosmology}
(World Scientific, Singapore, 1986).

\noindent\noindent\lbrack3] K. Halpern and K. Huang, Phys. Rev. Lett., 74,
3526 (1995); \textit{Phys. Rev}. \textbf{53}, 3252 (1996).

\noindent\lbrack4] V. Periwal, \textit{Mod. Phys. Lett}. \textbf{A 11}, 2915 (1996).

\noindent\noindent\lbrack5] Y. Choquet-Bruhat and J.W.York, in \textit{The
Cauchy Problem, General Relativity and Gravitation I}, A. Held, ed. (Plenum,
New York, 1980) p.99.

\noindent\lbrack6] A. G. Riess et al.,\textit{ Astrophys. J.} \textbf{659}, 98 (2007).

\noindent\lbrack7] R.P. Feynman, in \textit{Progress in Low Temperature
Physics, Vol.1}, C.J. Gorter, ed. (North-Holland, Amsterdam, 1955), p.17.

\noindent\lbrack8] K.W. Schwarz, \textit{Phys. Rev. Lett.} 49, 283 (1982);
\textit{Phys. Rev. }\textbf{B 38}, 2398 (1988).

\noindent\lbrack9] D. Kivotides, C.F. Barenghi, and D.C.
Samuels.\textit{\ Phys. Rev. Lett}. \textbf{87}, 155301 (2001).

\noindent\lbrack10] K. Huang and H.-F. Meng, \textit{Phys. Rev. Lett.},
\textbf{69}, 644 (1992).

\bigskip

{\LARGE Figure captions}

\noindent

Fig.1. Halpern-Huang potential $U_{b}$ for a real field $\left(  N=1\right)
$, with $b=1,c=0.1.$ Its argument is $z=8\left(  \pi a\phi\right)  ^{2}$.
Right panels show $\phi\left(  t\right)  ,$ and the Hubble parameter $H\left(
t\right)  $ in a log-log plot. The curvature parameter is $k=-1$. The initial
conditions are given by $\left\{  a_{0},\phi_{0},\dot{\phi}_{0}\right\}
=\left\{  1,.01,.1\right\}  .$ We obtain $p=0.81$. All quantities are measured
in Planck units, in which $4\pi G=1$.

Fig.2. The potential is the superposition $U=0.1U_{b=1}-0.2U_{b=2}$, with
$c=1.$ This makes $U>0$ near $\phi=0,$ and gives a "push" to the initial
acceleration of $a\left(  t\right)  .$ We set $k=0$, and the initial
conditions are $\left\{  a_{0,}\phi_{0},\dot{\phi}_{0}\right\}  $ $=\left\{
1,0,.1\right\}  .$ We obtain $p=0.075.$

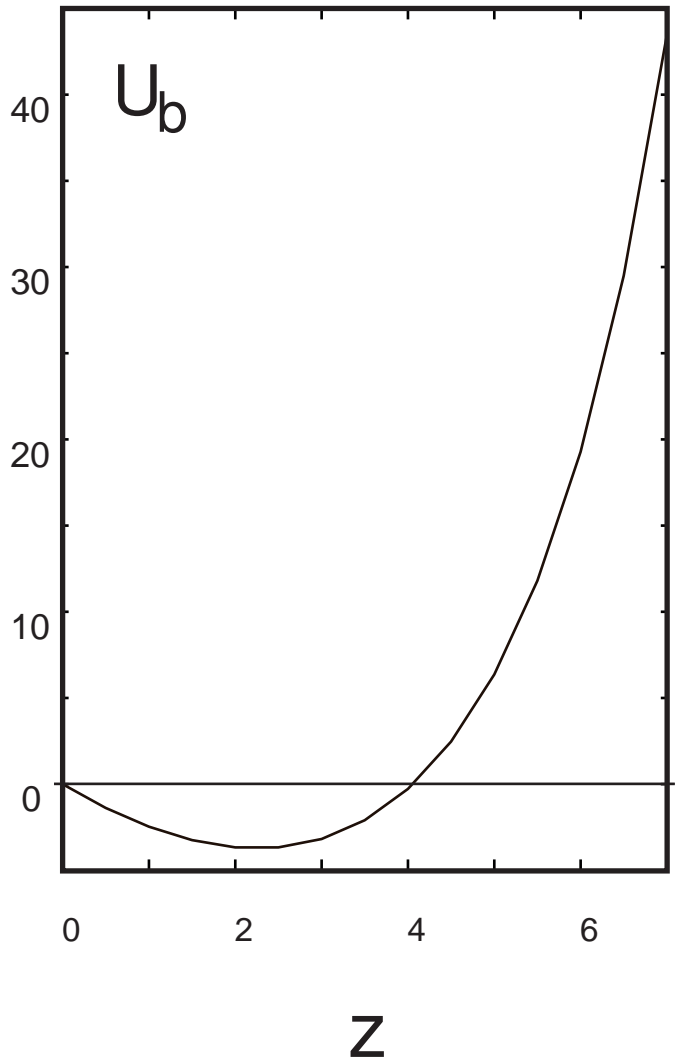
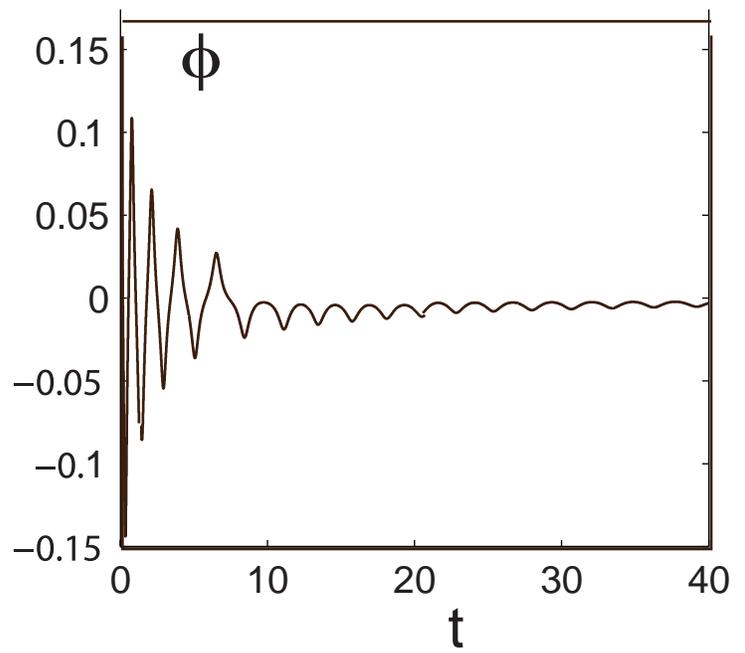
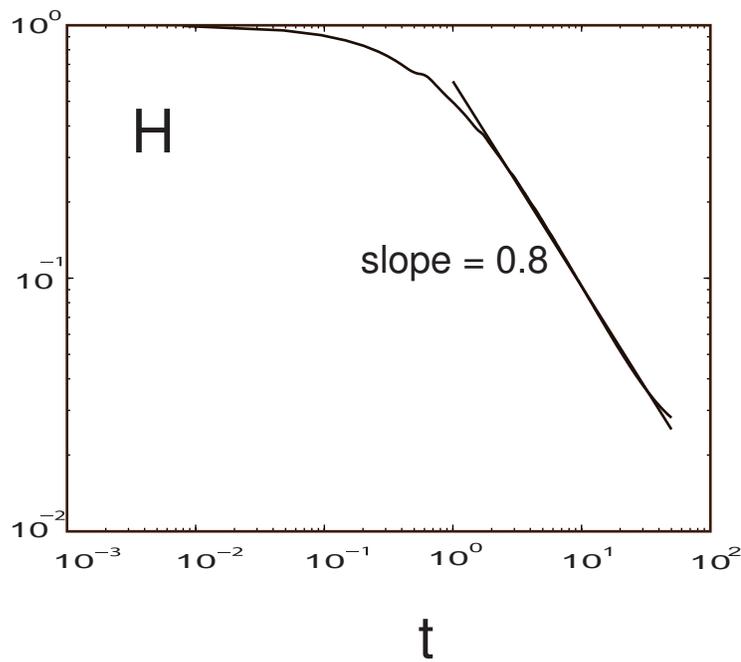
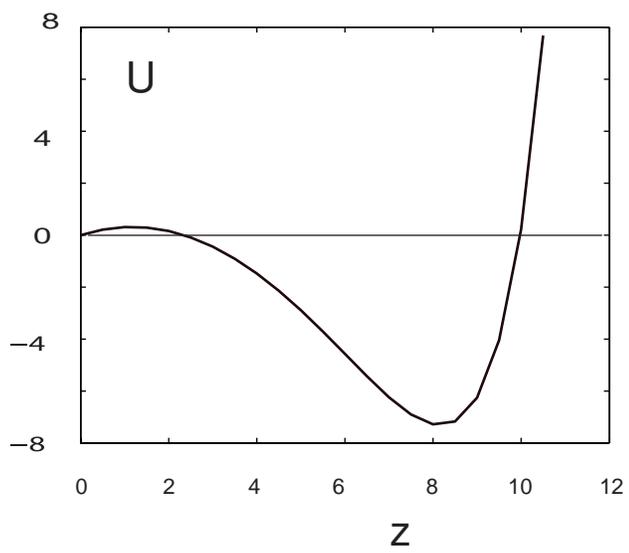
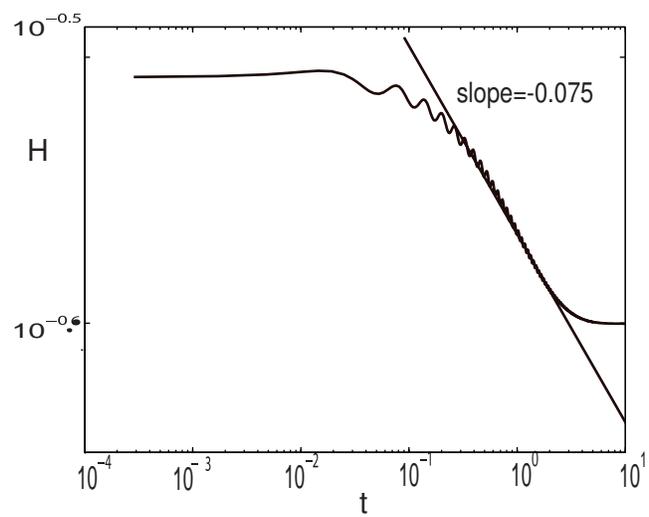
\end{document}